\documentstyle[12pt]{article}

\newlength{\extraspace}
\newlength{\extraspaces}
\newcommand{\be}{\begin{equation}
\addtolength{\abovedisplayskip}{\extraspaces}
\addtolength{\belowdisplayskip}{\extraspaces}
\addtolength{\abovedisplayshortskip}{\extraspace}
\addtolength{\belowdisplayshortskip}{\extraspace}}
\newcommand{\ee}{\end{equation}}

\newcommand{\bq}{\begin{eqnarray}
\addtolength{\abovedisplayskip}{\extraspaces}
\addtolength{\belowdisplayskip}{\extraspaces}
\addtolength{\abovedisplayshortskip}{\extraspace}
\addtolength{\belowdisplayshortskip}{\extraspace}}
\newcommand{\eq}{\end{eqnarray}}

\newcommand{\newsection}[1]{
\vspace{15mm}
\pagebreak[3]
\addtocounter{section}{1}
\setcounter{equation}{0}
\setcounter{subsection}{0}
\setcounter{footnote}{0}
\begin{flushleft}
{\large\bf \thesection. #1}
\end{flushleft}
\nopagebreak
\medskip
\nopagebreak}

\begin{document}
\hbox{}
\nopagebreak
\vspace{-3cm}
\addtolength{\baselineskip}{.8mm}
\baselineskip=24pt
\begin{flushright}
{\sc UMN-TH-1403-95}\\
{\sc TPI-MINN-95/23-T}\\
{\sc OUTP}- 95 - 30 P\\
hep-th@xxx/9507137\\
July  1995
\end{flushright}

\begin{center}
{\Large  The Two Phases of Topologically Massive
  Compact $U(1)$  Theory.}\\
\vspace{0.1in}

{\large Ian I. Kogan}
\footnote{ On  leave of absence
from ITEP,
 B.Cheremyshkinskaya 25,  Moscow, 117259, Russia}\\
{\it  Theoretical
 Physics,
1 Keble Road, Oxford, OX1 3NP, UK\footnote{permanent address}
 \\and\\
 Theoretical Physics Intstitute and Physics Department\\
  University of Minnesota,  Minneapolis, MN 55455, USA}\\
and\\
{\large Alex Kovner}\\
{\it Physics Department, University of Minnesota,
  Minneapolis, MN 55455, USA}\\
\vspace{0.1in}

\vspace{0.1in}

{\sc  Abstract}
\end{center}

\noindent
The mean field like gauge invariant variational method formulated
recently, is applied to a topologically massive QED in 3
dimensions. We find that the theory has a phase transition in the
Chern Simons coefficient $n$.  The phase transition is of the
Berezinsky-Kosterlitz - Thouless type, and is triggered by the liberation of
Polyakov monopoles, which for $n>8$ are tightly bound into pairs. In
our Hamiltonian approach this is seen as a similar behaviour of the
magnetic vortices, which are present in the ground state wave functional
of the compact theory.
For $n>8$, the low energy behavior of the theory is the same as in the
noncompact case.  For $n<8$ there are no propagating degrees of
freedom on distance scales larger than the ultraviolet cutoff.  The
distinguishing property of the $n<8$ phase, is that the magnetic flux
symmetry is spontaneoously broken.

\vfill

\newpage
\newsection{Introduction.}

In recent years topologically massive gauge theories,
(TMGT) i.e. three-dimensional gauge theories with a Chern-Simons
 term \cite{tmgt} have attracted a lot of attention.

In this paper we study
  the simplest model of this variety - the
 topologically massive $U(1)$ gauge theory.
 The Lagrangian of this theory in the naive continuum limit is
 \begin{equation}
L = \frac{1}{4g^2}F_{\mu\nu}F_{\mu\nu} +
\frac{n}{8\pi} \epsilon_{\mu\nu\lambda} F^{\mu\nu}A_{\lambda}
\label{tmgt}
\end{equation}
 where the charge $g^2$ has a dimension of mass and the
 Chern-Simons coefficient is a number. In a noncompact case
 this Lagrangian describes
  free massive particles
 with mass $2\kappa = n g^2/2\pi$.
 We will be interested in the
 compact case. Our motivation  to study this problem is twofold.

 First, compact three-dimensional QED without Chern-Simons
 term is confining due to Polyakov's monopoles-instantons \cite{polyakov}.
 The coexistence of these monopoles with Chern-Simons term is an
 interesting question. It was studied  to some extent in several
 papers \cite{CSmonopole} where it was  argued that the monopoles are
 irrelevant in the presence of the Chern-Simons term, the
 theory looses its confining properties and behaves in all aspects
 as a non-compact theory. On the other hands these arguments are
 not very rigorous and the question of the relevance of the monoples
is not completely settled.

 The second issue arises from the connection between the $2+1$
topologically massive gauge theory with compact $U(1)$ group and an
induced $2$ dimensional $XY$ model on the space - time boundary of the
2+1 dimensional manifold.  This model arises because the Chern-Simons
term in the TMGT Lagrangian (\ref{tmgt}) is not gauge invariant under
the gauge transformations which do not vanish on the boundary. These
gauge degrees of freedom become physical and have nontrivial dynamics
on the boundary.  One can
show that the induced two-dimensional action has the form

\begin{equation}
S_{XY} = \frac{n}{8\pi}\int d^2 x ~\partial_i\phi\partial_i\phi
\label{XYaction}
\end{equation}

 This is the action for the massless two-dimensional scalar field
  $\phi$ which is a parameter of a gauge transformation in the
  original three dimensional theory.  The question of whether the
  three dimensional theory is compact or not is of crucial
  importance. In the noncompact case the gauge parameter varies
  from $-\infty$ to $\infty$ and the theory on the boundary is the
  theory of free massless bosons - the conformal field theory with
  $c=1$. In the compact case with gauge group $U(1)$, the gauge
  parameter lives on a circle $S^1$ and eq.(\ref{XYaction}) defines
  the $XY$ model which has the famous Berezinsky-Kosterlitz-Thouless
  (BKT) phase transition \cite{bkt}.  The difference between the XY
  model and the free scalar field is the existence of singular field
  configurations (vortices). The $XY$ model has two phases.  At large $n$
  the vortices are bound into dipolar molecules and do not affect long
  distance physics.  This phase is described by the conformal $c=1$
  theory.  At small $n$ the vortices are in the plasma phase.  They
  dominate the statistical sum and lead to emergence of finite
  correlation length. The critical value of $n$ for the action
  (\ref{XYaction}) is $n=8$. Thus, due to the connection between the XY
  model and the compact $U(1)$ TMGT one is lead to think that the
  three-dimensional theory should also have two phases.  This phase
  transition and existence of a new phase in the topologically massive
  gauge theory as well as some related issues, for example $n
  \rightarrow 1/n$ duality in both theories ($R \rightarrow 1/R$ in a
  standard notation in string theory) have been discussed some time
  ago by one of the authors \cite{kogan}.  However the complete picture
  of the phase transition at small $n$ was not clear at that time.

  In this paper we study the  topologically massive compact $U(1)$
 gauge theory using the  gauge invariant variational approach
 developed in \cite{kk1}, \cite{kk2}.
In ref. \cite{kk2} this method was used to study compact QED$_3$
without the Chern Simons term.  These results are
in agreement with the well known scenario of confinement due to
magnetic monopoles. In the framework of the hamiltonian approach
of \cite{kk2} the monopoles appear as topologically nontrivial
configurations of the gauge group when the trial vacuum wave functional
is projected onto the gauge invariant subspace of the Hilbert space.
The advantage of this approach is a straightforward generalization
 in the case of non-zero Chern-Simons term. As opposed to that,
in  the
 path integral approach  in the presence of the
 non-zero Chern-Simons term, the monopoles are not solutions
 of the classical equations of motion anymore.
One has to consider complex-valued
 solutions of the equations of motion and their meaning is not
 completely clear \cite{CSmonopole}.

We show that the model has two phases depending on the value of the
 Chern-Simons coefficient $n$. The existence of these two phases is
 related to the different behaviour of the monopoles. At $n > 8$ the
 monopoles are irrelevant for the infrared physics and the theory
 behaves as in the noncompact case and induces conformal $c=1$ model
 on the boundary. This is also in a qualitative agreement with the
 results of ref. \cite{CSmonopole}.

 For $n <8$ the monopoles condense in a new ground state. In this phase
 they are important for the infrared physics and the induced theory on the
 boundary is a deformed $c =1$ model corresponding to the plasma phase.
 Let us note that even though there is a monopole condensate in this
 phase, there is no confinement for any nonzero $n$.

  The transition between these two phases indeed takes place precisely
 at $n=8$, which is a strong argument in support of the idea that all
 phenomena in induced two-dimensional dynamics have they counterparts
 in an original three-dimensional theory.

 The organisation of the paper is the following. In the next section
 we consider the Hamiltonian formalism for the topologically massive
 gauge theory. First we discuss the noncompact case and then modify it
 for the case of compact $U(1)$. We introduce the gauge invariant
 trial wave functional. In section 3 we discuss some properties of
 this wave functional and the interpretation of the calculations in
 terms of gases of magnetic and electric vortices.  In section 4 we
 calculate the average energy in this trial functional and minimize it
 with respect to our variational parameters. We will demonstrate that
 for large $n$ the long distance properties of the ground state are
 the same as in the noncompact case and then will investigate the case
 of small $n$.  In conclusion we discuss the properties of the new
 $n<8$ phase.

\section{The model and the setup.}
\subsection{The noncompact limit.}

Let us start with setting up the Hamiltonian formalism for the
theory.  To do that consider first a noncompact theory. It is described
by the following Hamiltonian
\begin{equation}
H=\frac{1}{2}[E_i^2+B^2]
\label{hn}
\end{equation}
augmented by the Gauss' law constraint
\begin{equation}
C(x)=\partial_i(E_i-2\kappa\epsilon_{ij}A_j)=0
\label{conn}
\end{equation}
The commutation relations satisfied by the fields are
\begin{equation}
[A_i(x),A_j(y)]=0;\ \ [E_i(x),A_j(y)]=i\delta_{ij}\delta(x-y);\ \
[E_i(x),E_j(y)]=2i\kappa\epsilon_{ij}\delta(x-y)
\label{comr}
\end{equation}
The electric field $E_i$ can be represented in terms of the momentum,
canonically
conjugate to the vector potential $A_i$ as
\begin{equation}
E_i=-\Pi_i+\kappa\epsilon_{ij}A_j
\end{equation}
This  Hamiltonian formulation follows directly from the
 Lagrangian  (\ref{tmgt}).

The Gauss' law operator $C(x)$ generates time independent gauge
transformation.  The elements of the gauge group are the operators
\begin{equation}
U_\phi=\exp\left\{i\int d^2x \phi(x)C(x)\right\}=
\exp\left\{i\int d^2x\partial_i\phi(\Pi_i(x)+
\kappa\epsilon_{ij}A_j(x))\right\}
\label{gt}
\end{equation}

It is straightforward to check, that both $E_i$ and $B$ commute with
the Gauss' law, and are therefore gauge invariant operators.  The
physical Hilbert space of the theory contains only states which are
invariant under the action of $U$
\begin{equation}
U_\phi|\Psi>=|\Psi>
\end{equation}
It is most convenient for our purposes to work in the field
basis. Then, defining
\begin{equation}
\Phi[A_i]=<A_i|\Phi>
\end{equation}
we find
\begin{equation}
<A|U_\phi|\Phi>=\exp\left\{i\kappa\int d^2x
\partial_i\phi(x)\epsilon_{ij}A_j(x)\right\}
\Phi[A_i+\partial_i\phi]
\end{equation}
A gauge invariant state can be then constructed from an arbitrary state
$\Phi$ by averaging over the gauge group
\begin{equation}
\Psi[A_i]=\int D\phi\exp\left\{i\kappa\int d^2x
 \partial_i\phi(x)\epsilon_{ij}
A_j(x)\right\}
\Phi[A_i+\partial_i\phi]
\label{gis1}
\end{equation}
In the case of noncompact theory the integral over $\phi$ can be
performed, and one can give a more compact representation of a gauge
invariant state as
\begin{equation}
\exp\{i\kappa\int \frac{\partial_i}{\partial^2}A_iB\}\Phi[B]
\end{equation}
However, we are going to deal with the compact theory, in which case the
representation eq.(\ref{gis1}) is more helpful.

\subsection{Compactifying the gauge group. The vortex operator.}

The next step in our discussion is to compactify the gauge group. As
discussed in some detail in \cite{kk2} this means, that we have to
enlarge the gauge group so that it includes operators which create
pointlike magnetic vorices of integer strength. These operators
satisfy the following commutation relation
\begin{equation}
V_k^\dagger (x)B(y)V_k(x)=B(y)+\frac{2\pi k}{g}\delta^2(x-y)
\label{commut}
\end{equation}
with $g$ - a dimensional constant, determining the radius of
compactness and $k$ - arbitrary integer.  Naively this just means
including in the gauge group the gauge transformations of the form
eq.(\ref{gt}), corresponding to singular functions
\begin{equation}
\phi(x)=\frac{n}{g}\theta(x-x_0)
\end{equation}
where $\theta$ - is the planar angle, and with the understanding that
the derivative of $\phi$ in eq. (\ref{gt}) should be
taken
modulo $2\pi$.
For future convenience let us introduce a special symbol to denote this
derivative
\begin{equation}
\Delta_i \theta(x)=\partial_i[\theta(x)]_{mod
2\pi}=\frac{\epsilon_{ij}x_j}{x^2}
\end{equation}
That is, these derivatives do not feel quantized discontinuities
in $\phi(x)$.

In other words, to compactify the theory we must limit physical
Hilbert space to states which are eigenstates of the operators
\begin{equation}
V(x)=\exp\left\{\frac{i}{g}\int d^2y\Delta_i\theta(x-y)
[E_i(y)-2\kappa\epsilon_{ij}A_j]\right\}
\label{vortex1}
\end{equation}
with a unit eigenvalue.  For $\kappa=0$ this procedure is
straightforward and was described in \cite{kk2}. However, for a
nonzero $\kappa$ one should be a little more careful.  The point is,
that the operators defined in eq. (\ref{vortex1}) do not commute with
noncompact gauge transformations, and also do not commute between
themselves, but rather satisfy
\begin{equation}
V(x)V(y)=V(y)V(x)e^{i\kappa\alpha(x,y)}
\end{equation}
where $\alpha(x-y)$ is a c-number. Obviously, if $\kappa\alpha\ne 2\pi
m$, one can not impose the condition $V(x)|\Psi>=|\Psi>$
simultaneously for all points $x$.  Fortunately, eq. (\ref{commut}) does not
define the operator $V$ completely, and we can take advantage of this
freedom.  Let us consider the following operator
\begin{equation}
V_C(x)=\exp\left\{\frac{i}{g}\int d^2y[\Delta_i\theta(x-y)
E_i(y)-2\kappa\epsilon_{ij}\partial_i\theta(x-y)A_j]\right\}
\label{vortex}
\end{equation}
Note, that the function multiplying $A_j$ in the exponential is now a
{\it bona fide} derivative. Since the planar angle is defined relative
to some direction, the function $\theta$ has a discontinuity of $2\pi$
along some curve $C$, which starts at the origin and goes to
infinity. Its derivative therefore has the form
\begin{equation}
\partial_i\theta(x)=\Delta_i\theta(x)+2\pi\epsilon_{ij}\hat n^C_j(x)\delta(x-C)
\end{equation}
where $\hat n^C_i(x)$ is the unit tangent vector to the curve $C$ at the
point $x$.  The operator $V(x)$ defined in eq. (\ref{vortex}) still
satisfies eq. (\ref{commut}), but now is invariant under the noncompact
part of the gauge group\footnote{To see this, note that the $A$ -
dependent term in the exponential, integrating by parts can be
rewritten as $\int \phi B$, which is explicitly gauge invariant.  In
principle one could worry about the surface term, which we omitted
$\oint_{x^2\rightarrow\infty} ds^i\phi\epsilon_{ij} A_j$. In the
present case however, it is harmless for two reasons. First, because
the vector potential itself is massive, and therefore vanishes at
infinity; and second, since for all practical purposes it is enough to
consider transformations with nonzero number of vortices and
antivortices, but with net zero vorticity, and for those the function
$\phi$ itself vanishes at infinity.}, and satisfies
\begin{equation}
V_{C_1}(x)V_{C_2}(y)=V_{C_2}(y)V_{C_1}(x)e^{i\frac{8\pi^2\kappa}{g^2}N(C_1C_2)}
\end{equation}
where $N(C_1C_2)$ is the number of intersections between the curves
$C_1$ and $C_2$.  It is clear therefore, that if we choose the radius
of compactness so that
\begin{equation}
4\pi\kappa=ng^2
\label{quant}
\end{equation}
with arbitrary integer $n$, the operators $V_C(x)$ commute at all
points and for arbitrary choice of the set of curves $C$.  Only in
this case the compact theory will not depend on the choice of the
curves, and will therefore be rotationally and translationally
invariant.  We see, therefore that for nonzero $\kappa$ the radius of
campactness must be quantized according to eq. (\ref{quant}). This is, of
course in complete agreement with the fact that in nonabelian Yang
Mills theories, where the gauge group is necessarily compact, the
ratio of the coupling constant and the Chern - Simons coefficient is
quantized in the same way.

The expression for the vortex operator $V$ is simpler on a subspace of
the Hilbert space invariant under the noncompact part of the gauge
group. On these states the Gauss' law is implemented
exactly. Integrating the second term in the exponential in
eq. (\ref{vortex}) by parts and using the Gauss' law, we get
\begin{equation}
V_s=\exp\{i\int s_i(y)E_i(y)\}
\label{V}
\end{equation}
with $s_i(x)= 2\pi/g\epsilon_{ij}\hat n^C_j(x)\delta(x-C)$.  We find it
more convenient to label the operator $V$ by the vector function $s_i$
rather than by the curve $C$. They are of course in one to one
correspondence.

There is one subtlety related to the algebra of the
operators $V_s$ for odd $n$. Although the vortex operators commute for
any $n$, their group multiplication properties are not completely
trivial.
\begin{equation}
V_{s_1}V_{s_2}=e^{i\pi n N(1,2)}V_{s_1+s_2}
\label{prod}
\end{equation}
where N(1,2) is again the intersection number of the two curves.  For
even $n$ the extra phase factor is always an integer of $2\pi$. In
this case one can construct physical states by requiring
$V_s\Psi=\Psi$ for all $s$. For odd $n$ the phase factor can be an odd
integer of $\pi$.  We therefore can not require that the wave
functional be invariant under the action of all $V_s$. This however
can be remedied in the following way.  Let us define for future
convenience the densities $\rho$ and $\sigma$ by
\begin{equation}
\frac{g}{2\pi}\epsilon_{ij}\partial_is_j=\rho,\ \ \
\frac{g}{2\pi}\partial_is_i=\sigma
\label{rhosigma}
\end{equation}
With our definition of $s_i$, both $\rho$ and $\sigma$ satisfy
\begin{equation}
\int_S d^2x \rho(x)=2\pi n_1,
\int_S d^2x \sigma(x)=2\pi n_2
\label{dense}
\end{equation}
with integer $n_1$ and $n_2$, for any finite area $S$.
Then the field $\tau$
\begin{equation}
\partial^2\tau=2\pi\sigma
\label{tau}
\end{equation}
takes integer values.

Now, note that the phase factor in eq. (\ref{prod})
can be represented as
\begin{equation}
\exp\{i\pi n N(1,2)\}=\exp\{i\kappa\int \epsilon_{ij}s_i^1s_j^2\}=
\exp\{i\frac{2\pi \kappa}{g^2}\int [\tau^1\rho^2-\rho^1\tau^2]\}
\end{equation}
Due to the property eq. (\ref{dense}) this can be further rewritten as
\begin{equation}
\exp\{i\frac{n}{2}\int [\tau^1\rho^2+\rho^1\tau^2]\}=
\exp\{i\frac{n}{2}\int [(\tau^1+\tau^2)(\rho^1+\rho^2)]\}
\end{equation}
It is clear now, that a modified vortex operator, defined by
\begin{equation}
V_s=\exp\{i\int s_iE_i+i\frac{n}{2}\int\tau(s)\rho(s)\}
\end{equation}
has a simple group multiplication rule
\begin{equation}
V_{s_1}V_{s_2}=V_{s_1+s_2}
\end{equation}
One therefore can require consistently, that these operators for all
$s_i$ leave physical states invariant. In the following we will,
however disregard this subtlety and for simplicity use the vortex
operators defined in eq. (\ref{V}). Thus our calculations in the way they
are presented below are valid only for even $n$.

Before we continue, we wish to remark that the densities $\rho$ and
$\sigma$ as defined in eq. (\ref{dense}) have a very clear physical
meaning.  The vortex operator $V_s$ has the following commutation
relations with electric and magnetic fields
\begin{eqnarray}
V^\dagger_s(x)B(y)V_s(x)&=&B(y)+\frac {2\pi}{g}\rho(y)\\ \nonumber
V^\dagger_s(x)\partial_iE_i(y)V_s(x)&=&\partial_iE_i(y)+ng\rho(y)\\
\nonumber
V^\dagger_s(x)\epsilon_{ij}\partial_iE_j(y)V_s(x)&=&\epsilon_{ij}\partial_i
E_j(y)+ng\sigma(y)
\end{eqnarray}
Therefore the operator $V_s$ when acting on any state creates magnetic
vortices (with magnetic flux quantized in units of $2\pi/g$) with the
density $\rho$, and also creates electric vortices (with vorticity
quantized in units of $ng$) with the density $\sigma$.

\subsection{The Hamiltonian.}

As just noted, the operators $B$ and $E_i$ do not commute with
$V$. The Hamiltonian of the noncompact theory should, however be
invariant under the complete compact group. It must therefore be
slightly modified from the simple form of eq. (\ref{hn}).  The most
natural way to do it, is to use gauge invariant operators which in the
naive continuum limit reduce to the standard $B^2$ and $E^2$ terms in
the continuum hamiltonian. For the magnetic part of the energy density
we take therefore
\begin{equation}
H_B(x)=\frac{a^{-4}}{g^2m^2}[1-{\rm Re}e^{igma^2B(x)}]
\label{HB}
\end{equation}
Here $a$ is the ultraviolet regulator which has a dimension of
distance, and has the meaning of the lattice spacing.\footnote{In
general, in the following wherever ultraviolet regularization is
needed we assume lattice regularization with lattice spacing
$a$. Accordingly, all derivatives should be understood as symmetric
lattice derivatives etc.} The constant $m$ is an arbitrary integer,
$m<n$.

As opposed to the case $\kappa=0$, the electric part of the
Hamiltonian should also be modified. Again, guided by the naive
continuum limit and gauge invariance, we write for the electric part
\begin{equation}
H_E(x)=\frac{g^2l^2a^{-2}}{2\pi}\sum\limits_{i=1,2}
[1-{\rm Re}e^{\frac{i}{l}\alpha_i(x)E_i(x)}]
\label{HE}
\end{equation}
The vector function $\alpha_i(x)$ is defined on the links of the lattice
in terms of the unit vector $\hat n_i(x)$
parallel
to the given link as
\begin{equation}
\alpha_i(x)=\frac{2\pi a}{g}\epsilon_{ij}\hat n_j(x)
\end{equation}
In order for $H_E$ to commute with the vortex operator $V$, the
integer $l$ must be a divisor of $n$: $n/l=integer$.  Note, that for
$l=1$, the operator $H_E(x)$ is equal to the sum of two vortex
operators. In that case, it is trivial on all physical states.  The
same is true for $H_B$ for $m=n$. In that case, using the Gauss' law
one can see that $H_B$ is given entirely in terms of a vortex operator
$V_s$ where the function $s_i$ corresponds to the contour $C$ which is
the boundary of the elementary plaquette on the lattice.  Therefore it
is obvious, that in order for the dynamics of the theory defined with
this Hamiltonian to be the most nontrivial, one should choose the
largest possible $l$ and the smallest possible $m$.  However, it turns
out, that to avoid subtleties similar to the ones encountered for odd
$n$, it is convenient to choose both $m$ and $n/l$ to be even
numbers. We will therefore take $m=2$ and $l=n/2$ in the rest of this
paper.

\section{The Variational wave functional.}

Now it is straightforward to write down a general wave functional,
which is invariant under the whole compact gauge group\footnote{The
last, c-number, term in the phase factor appears, since the two
components of the electric field do not commute, and is readily
obtained using Baker-Campbell-Hausdorff formula.}
\begin{equation}
\Psi[A_i]=\int Ds_iD\phi\exp\left\{i\kappa
\int d^2x \left[\epsilon_{ij}(s_i(x)-\partial_i\phi(x))A_j(x)
+\epsilon_{ij}\partial_i\phi(x)s_j(x)\right]\right\}
\Phi[A_i-\partial_i\phi-s_i]
\label{gis}
\end{equation}

Our purpose is to find a wave functional of the vacuum of this theory
using a gauge invariant generalization of a gaussian variational
approximation. We shall minimize the VEV of the energy on the
following set of explicitly gauge invariant states:
\begin{equation}
\Psi[A_i]=
\int Ds_iD\phi\exp\left\{i\kappa
[s_i\epsilon_{ij}A_j-\partial_i\phi
\epsilon_{ij}A_j
+\epsilon_{ij}\partial_i\phi s_j]
-\frac{1}{2} A_i^{\phi,s}G^{-1}A_i^{\phi,s}
\right\}
\label{wf}
\end{equation}
For convenience we have introduced the following notation
\begin{equation}
A_i^{\phi(x),s(x)}= A_i(x)-\partial_i\phi(x)-s_i(x)
\end{equation}
and have switched to the matrix notations, so that
\begin{equation}
a_iMb_i=\int d^2xd^2ya_i(x)M(x-y)b_i(y)
\end{equation}

In the next section we are going to calculate the expectation value of
the Hamiltonian on this set of states, and functionally minimize it
with respect to $G(x-y)$, which is our variational parameter.

\subsection{ The noncompact limit.}
We start with solving the noncompact limit of the theory. This serves
to illustrate the method, and also to get a clear idea of what to
expect of the variational function $G(x)$.  Our variational state now
is given by eq.(\ref{wf}), but without the integration over $s_i$.
The Hamiltonian is the simple quadratic Hamiltonian of eq.(\ref{hn}).
The exact ground state wavefunctional in this case is therefore
gaussian. So our variational calculation in this case will give the
exact result.
The expectation value of any gauge invariant operator is given by
the following path integral
\begin{equation}
<O[A]>=Z=\int D\phi DA_i
\exp\left\{-\frac{1}{2} A_i^{\phi}G^{-1}A_i^{\phi}+ i\kappa
 \partial_i\phi)\epsilon_{ij}A_j
\right\}O[A]
\exp\left\{-\frac{1}{2} A_iG^{-1}A_i\right\}
\end{equation}
The integrals over $A_i$ and $\phi$ are both Gaussian, and are
easily performed. Calculating the relevant averages we obtain
\begin{equation}
<B^2(x)>=\frac{1}{2}\int \frac{d^2p} {(2\pi)^2}p^2[G^{-1}(p)+\kappa^2G(p)]^{-1}
\end{equation}
\begin{equation}
<E^2(x)>=\frac{1}{2}\int \frac{d^2p} {(2\pi)^2}\left[G^{-1}(p)+\kappa^2G(p)+
4\kappa^2[G^{-1}(p)+\kappa^2G(p)]^{-1}\right]
\end{equation}
Here $G(k)$ is the Fourier transform of the variational function
$G(x)$.

The expectation value of the energy is
\begin{equation}
2V^{-1}<H>=\frac{1}{2}\int \frac{d^2p} {(2\pi)^2}\left[G^{-1}(p)+\kappa^2G(p)+
(p^2+4\kappa^2)[G^{-1}(p)+\kappa^2G(p)]^{-1}\right]
\end{equation}
Minimizing this expression with respect to $G(p)$ we obtain
\begin{equation}
G^{-1}(p)+\kappa^2G(p)=\sqrt{p^2+4\kappa^2}
\label{ncsol}
\end{equation}
As is clear from this calculation, the function
\begin{equation}
D(p)=[G^{-1}(p)+\kappa^2G(p)]^{-1}
\label{D}
\end{equation}
plays the role of the propagator. For example, the equal time
propagator of magnetic field
is given by
\begin{equation}
<B(x)B(y)>=\frac{1}{2}\int \frac{d^2p}{(2\pi)^2}e^{ip(x-y)}
{(2\pi)^2}p^2[G^{-1}(p)+\kappa^2G(p)]^{-1}
\end{equation}
As expected, $D(p)$ in the noncompact case comes out as the propagator of
a free massive particle with mass $m^2=4\kappa^2$.

\subsection{The norm of the state in the compact theory.}

As a preamble to calculating the energy expectation
value, let us find the norm of the state eq. (\ref{wf}).
\begin{equation}
Z=\int Ds_i D\phi DA_i
\exp\left\{-\frac{1}{2}[ A_iG^{-1}A_i+
 A_i^{\phi,s}G^{-1}A_i^{\phi,s}]- i\kappa
 [(s_i-\partial_i\phi)\epsilon_{ij}A_j+\epsilon_{ij}\partial_i s_j]
\right\}
\label{exp1}
\end{equation}
The integrals over $A_i$ and over the noncompact part of the gauge
group, $\phi$ are gaussian. After performing them we are left with the
integral over the descrete variable $s_i$.
\begin{equation}
Z=Z_\alpha Z_\phi Z_s
\end{equation}
with
\begin{eqnarray}
Z_a&=&\det (\pi G)\\ \nonumber
Z_\phi&=&\int D\phi \exp\left\{
-\frac{1}{4}\int \partial_i\phi
D^{-1}\partial_i\phi\right\}=
\det\left[\frac{\partial^2}{4\pi}D^{-1}
\right]^{-1/2}\\
\nonumber
Z_s&=&\int Ds_i\exp\left\{-\frac{1}{4}\int \frac{d^2p}{(2\pi)^2}
\left[s_i(p)[\delta_{ij}-\frac{p_ip_j}{p^2}]
D^{-1}(p)s_j(-p)+\kappa^2s_i(p)\frac{p_ip_j}{p^2}
D(p)s_j(-p)\right]\right\}
\end{eqnarray}
where $D(p)$ is defined in eq.(\ref{D}).

A more convenient and physically intuitive representation of the
integral over $s_i$ can be given in terms of the variables $\rho$ and
$\sigma$ defined in eq.(\ref{rhosigma}).
\begin{equation}
Z_s=Z_\rho Z_\sigma
\end{equation}
with
\begin{equation}
Z_\rho  = \int D\rho
\exp\left\{-\frac{\pi^2}{g^2}\int \frac{d^2p}{(2\pi)^2}
\rho(p)\frac{1}{p^2}D^{-1}(p)\rho(-p)\right\}
\end{equation}
\begin{equation}
Z_\sigma  =  \int D\sigma
 \exp\left\{-\pi n\kappa \int \frac{d^2p}{(2\pi)^2}
\sigma(p)\frac{1}{p^2}D(p)\sigma(-p)\right\}
\end{equation}

One should keep in mind, that the measures $D\rho$ and $D\sigma$ are
not the same as the measure for the functional integration
over a free field. Rather, due to
eq.(\ref{dense}) both these path integrals are equivalent to
statistical sums for classical interacting gases:
\begin{eqnarray}
Z_\rho &=&\sum_{n_+,n_- = 0}^{\infty}
\prod_{\alpha=1}^{n_{+}}\prod_{\beta=1}^{n_{-}}
(a^{-2})^{n_++n_-}
dx_\alpha dx_\beta\\
\nonumber
&\exp&\left\{-[\sum_{\alpha,\alpha'}u(x_\alpha-x_{\alpha'})
+\sum_{\beta,\beta'}u(x_\beta-x_{\beta'})-
\sum_{\alpha,\beta}u(x_\alpha-x_{\beta})]
+\frac{(n_++n_-)}{2}u(0)
\right\} \nonumber
\label{rgas}
\end{eqnarray}

\begin{eqnarray}
Z_\sigma &=&\sum_{n_+,n_- = 0}^{\infty}
\prod_{\alpha=1}^{n_{+}}\prod_{\beta=1}^{n_{-}}
(a^{-2})^{n_++n_-}
dx_\alpha dx_\beta\\ \nonumber
&\exp&\left\{-[\sum_{\alpha,\alpha'}v(x_\alpha-x_{\alpha'})
+\sum_{\beta,\beta'}v(x_\beta-x_{\beta'})-
\sum_{\alpha,\beta}v(x_\alpha-x_{\beta})] +\frac{(n_++n_-)}{2}v(0)\right\}
\label{sgas}
\end{eqnarray}

The interparticle interaction potentials $u$ and $v$ are given by
\begin{eqnarray}
u(x)&=&\frac{2\pi^2}{g^2}\int \frac{d^{2}p}{(2\pi)^{2}}\frac{1}{p^2}
D^{-1}(p)
\cos(px)\\ \nonumber
v(x)&=&2\pi n\kappa
\int \frac{d^{2}p}{(2\pi)^{2}}\frac{1}{p^2}D(p)
\cos(px)
\label{potential}
\end{eqnarray}
Since $u(0)$ and $v(0)$ are singular, the last terms in the
exponential in the equations (\ref{rgas},\ref{sgas}) should be
understood, as usual in the regularized sense, that is at finite UV
cutoff: $u(0)$ ($v(0)$) should be substituted by $u(x=a)$ ($v(x=a)$).

Remembering the interpretation of $\rho$ and $\sigma$, discussed in
the previous section, we shall refer to the statistical mechanical
systems defined by eqs.(\ref{rgas}) and (\ref{sgas}) as gases of
magnetic vortices and electric vortices respectively.

The calculation of any expectation value (such as the expectation value of the
energy) in our trial wave functional should now proceed in the standard way.
The integration over the vector potential $A_i$ and the noncompact part of the
gauge group is always Gaussian and therefore trivial. After performing this
integral we will be invariably faced with the problem of calculating certain
correlators in the two vortex ensembles, eq. (\ref{rgas}, \ref{sgas}).
It is therefore worthwhile to try and understand some general features
of these ensembles.

First note, that the cases $\kappa=0$ and $\kappa\ne 0$ are essentially
different.  For $\kappa=0$ (or equivalently $n=0$), the
electric vortex partition function becomes completely trivial - the
interaction potential vanishes. This is of course to be expected,
since we know that the electric vortices do not play any role
in the theory without the Chern - Simons term \cite{kk2}.

Another crucial difference is that the behaviour of the magnetic
vortex interaction at large distances is significantly changed.  At
$n=0$ the function $D^{-1}(p)=G^{-1}(p)$ for the best variational
state vanishes at zero momentum \cite{kk2}.  Consequently, the
interaction potential between the magnetic vortices is short
range. For $\kappa\ne 0$ on the other hand this cannot happen, since
the expression $D^{-1}(p)=G^{-1}(p)+\kappa^2G(p)$ is bounded from
below by $2\kappa$. Therefore the interaction between the vortices is
logarithmic at large distances, and whatever $G$ is, we are dealing
with the Coulomb gas.  Obviously, for large $\kappa$ the gas will be
in the molecular phase, and one expects that it will have no effect on
the large distance physics. This is in agreement with the
 general  arguments of  ref.\cite{CSmonopole}, that monopoles
should be irrelevant in a Chern - Simons theory. However at smaller
$\kappa$ the Coulomb gas will be in the plasma phase, and will
certainly affect physics.  The effective temperature of the Coulomb
gas of magnetic vortices in our case is
$T_m = 2g^2 /2\kappa = 4\pi/n$, and one therefore expects
this change in behaviour to set in at $n=8$.

For nonzero $n$ the electric vortices behave in a similar way.  Taking
the noncompact expression $D^{-1}(p=0)=2\kappa$, we find that at large
distances the interaction between the electric vortices is also
logarithmic, and in fact has the same asymptotics as the magnetic
vortex potential. The effective temperature is also
$T_e= 4\pi/n $ and one again expects the
Berezinsky-Kosterlitz-Thouless like phase transition at $n=8$.

\subsection{The magnetic vortex gas.}

In the following we will need to calculate correlation functions of
the vortex densities. To facilitate this we use the standard trick
\cite{polyakov}, \cite{samuel} to rewrite the partition function
of a classical gas
in terms of a path integral over a scalar field.
First consider the magnetic vortices.
The exponential factor in eq. (\ref{rgas}) can be rewritten as
\begin{equation}
a^{-2(n_++n_-)}\int D\chi\exp\{-\frac{g^2}{4\pi^2}\chi p^2
D\chi+i\rho\chi\}
\end{equation}
The summation over the number of vortices (or integral over $\rho$)
can be performed exactly

\begin{equation}
\int D\rho \exp\{i\rho(x)\chi(x)\}=\delta(\exp\{i\chi(x)\}-1)
\end{equation}
This gives
\begin{eqnarray}
Z_\rho &=&\int D\chi\exp\{-\frac{g^2}{4\pi^2}\chi p^2
D\chi\}\Pi_x\delta(\exp\{i\chi(x)\}-1\} \\ \nonumber
&=& \int D\chi
D\alpha\exp\left\{-\frac{g^2}{4\pi^2}\chi p^2 D\chi +i\int_x
2\alpha(x)(\cos\chi(x)-1)\right\}
\end{eqnarray}
In general the integration over the lagrange multiplier field
$\alpha$ is nontrivial. However, when the density of vortices is
small, one can perform the integration over $\rho$ in the dilute gas
approximation, that is sum only over those configurations which have
only at most one vortex or antivortex in every point in space. This
is equivalent to substitute for $\alpha$ a constant
\begin{equation}
Z_\rho =
\int D\chi \exp\left\{-\frac{g^2}{4\pi^2}\chi p^2
D\chi+2a^{-2}\cos\chi(x)\right\}
\label{sine}
\end{equation}

To calculate the correlator of $\rho$ in the dilute gas approximation
one can add $i\rho J$ to the
vortex free energy, and calculate functional derivatives of the
resulting partition function with respect to $J$ at zero $J$. A simple
derivation gives
\begin{equation}
<\rho(x)\rho(y)>=u^{-1}(x-y)-<u^{-1}\chi(x)u^{-1}\chi(y)>
\end{equation}

The propagator of $\chi$ is easily calculated. At weak coupling
the interaction in eq. ({\ref{sine}) is very small ($<\cos \chi><<1$).
To first order in the interaction the only
contribution to the propagator comes from the tadpole diagrams. This
is easily seen by rewriting the cosine potential in equation
(\ref{sine}) in the normal ordered form
\begin{equation}
\cos \chi=<\cos \chi>:\cos \chi:=z a^2:\cos \chi:
\label{normalorder}
\end{equation}
One should be a little careful in the definition of the normal
ordering.  Ordinarily the normal odering would be performed relative
to the free theory with the propagator $u(x-y)$. In the present case
however the free propagator at small momentum behaves
like a propagator of a massless particle
$u(k)\rightarrow_{k\rightarrow 0}k^2$, and the bubble integral which
enters the calculation of $<\cos\chi>$ is infrared divergent.  This
problem can be overcome by performing normal ordering relative to a
massive theory.  This can be done selfconsistently, by including the
quadratic term in $:cos(\chi):$ into the free propagator.  In this
approximation the propagator of $\chi$ is
\begin{eqnarray}
\int d^2 x e^{ikx}<\chi(x)\chi(0)>=
\frac{1}{u^{-1}(k)+2z}
\end{eqnarray}
with $z$ determined selfconsistently by
\begin{eqnarray}
z=a^{-2}<cos(\chi)>=a^{-2}\exp\{-\frac{1}{2}<\chi^2>\}=
\\ \nonumber
a^{-2}\exp\left\{-\frac{1}{2}\int \frac{d^2p}{(2\pi)^2}
\left[\frac{g^2}{2\pi^2}p^2D(p)+2z\right]^{-1}\right\}
\label{fug}
\end{eqnarray}

The correlator of the vortex densities is then
\begin{equation}
K(k)=\int d^2x e^{ikx}<\rho(x)  \rho(0)>=\frac{2z}{1+2zu(k)}
\label{rhoc}
\end{equation}.

The existance of the critical point $n=8$ is straightforward to see in this
approximation. Using the result for the noncompact theory $D(0)=1/2\kappa$,
and anticipating the fact that the infrared assymptotics of the propagator
is the same in the noncompact theory, we can rewrite eq. (\ref{fug}) as
\begin{equation}
 z=a^{-2}(za^2e^{-c})^{n/8}e^{-\mu_\rho}
\label{z}
\end{equation}
Here the constant $c$ is the chemical potential of the pure Coulomb
gas. It's exact value is not important, but one has to remember that
it is of order one and positive. The chemical potential $\mu_\rho$
measures how different the magnetic vortex gas is from the Coulomb gas
and is defined by
\begin{equation}
\mu_\rho=\frac{1}{2}\int \frac{d^2p}{(2\pi)^2}
\left[\left(\frac{g^2}{2\pi^2}p^2D(p)+2z\right)^{-1}-\left(\frac{1}{\pi n}p^2
+2z\right)^{-1}\right]
\label{fuga}
\end{equation}
The chemical potential $\mu_\rho$ depends on $z$ very weakly for small
$z$.  Also, at weak coupling the chemical potential is very
big. Taking, for orientation the noncompact result for $D(p)$, we find
that $\mu_\rho\propto \frac {a^{-1}}{g^2}$. Under these circumstances
it is easily seen that for $n<8$ the solution for eq. (\ref{fuga})
exists, and is indeed parametrically much smaller than the ultraviolet
cutoff $a^{-2}$.
\begin{equation}
z\propto a^{-2}e^{-\frac{8}{8-n}\mu_\rho}
\label{z1}
\end{equation}

For $n>8$ one can check that the selfconsistency equation
eq. (\ref{fug}) has only the trivial solution $z=0$\footnote{ Even
though eq. (\ref{z1}) for $n>8$ seems to give $z\propto O(a^{-2})$ this
is not correct, since this equation was derived under assumption that
$z$ is small.}.

Importantly, the use of the noncompact expression for $D(p)$ is not at
all crucial for this derivation. Remember, that quite generally
$D(p)<1/2\kappa$.  The chemical potential for the magnetic vortices is
therefore always larger than the chemical potential in the Coulomb gas
that corresponds to the XY model. The dilute gas approximation is
known to work reasonably well for the XY model \cite{bkt}, and
we are assured therefore that the dilute vortex gas approximation for
the magnetic vortex gas should be reliable for any variational
function $G(p)$.

\subsection{ The electric vortex gas.}

While the magnetic vortex gas can be consistently treated in the
dilute gas approximation, the behaviour of the electric vortex gas is
more complicated.  A dilute electric gas approximation is easily
formulated along the same lines.
This results in a partition function:
\begin{equation}
Z_\sigma =
\int D\psi \exp\left\{-\frac{1}{4\pi n\kappa}\psi p^2
D^{-1}\psi+\int d^2x 2a^{-2}\cos\psi(x)\right\}
\label{psi}
\end{equation}
The ``bubble summation'' then gives the propagator of the field $\psi$
\begin{equation}
\int d^2 x e^{ikx}<\psi(x)\psi(0)>=
\frac{1}{v^{-1}(k)+2\zeta}
\end{equation}
with $\zeta$ determined selfconsistently by
\begin{equation}
\zeta=a^{-2}<cos(\psi)>=
a^{-2}\exp\left\{-\frac{1}{2}\int \frac{d^2p}{(2\pi)^2}
\left[\frac{1}{2\pi n\kappa}p^2D^{-1}(p)+2\zeta\right]^{-1}\right\}
\label{zeta}
\end{equation}
This can be recast into the form
\begin{equation}
 \zeta=a^{-2}(\zeta a^2e^{-c})^{n/8}e^{-\mu_\sigma}
\label{zet}
\end{equation}
with the chemical potential $\mu_\sigma$
\begin{equation}
\mu_\sigma=\frac{1}{2}\int \frac{d^2p}{(2\pi)^2}
\left[\left(\frac{1}{2\pi
n\kappa}p^2D^{-1}(p)+2\zeta\right)^{-1}-\left(\frac{1}{\pi n}p^2
+2\zeta\right)^{-1}\right]
\label{fugb}
\end{equation}

One should however ask oneself, under what circumstances is the dilute
gas approximation valid. A necessary condition for this is that the
interaction energy between the vortices is not negligibly small
even at very short distances of order of the ultraviolet cutoff.
It is clear therefore, that the diluteness of the electric vortex gas
holds only for some restricted set of the variational functions
$D$. Let us take, for example the noncompact form for $D(p)$, so that
for small momenta $D(p)=1/2\kappa$ and in the ultraviolet region
$D(p)\propto (p^2)^{-1/2}$. In this case the interaction potential
between the vortices is logarithmic at large distances ($x>1/\kappa$)
but at short distances is very weak $v(x)\propto n\kappa x$.  The
vortex gas will therefore be dilute on all distance scales only if
$n>(\kappa a)^{-1}$, that is either for very large values of $n$, or
in a theory where the mass of the photon is close to the ultraviolet
cutoff.  The former case corresponds to extremely weak coupling, while
the latter to the limit of pure Chern - Simons theory without the
Maxwell term.  On the other hand for finite $n$ and $\kappa
a\rightarrow 0$ (naive continuum limit), the electric vortices
therefore hardly feel the presence of each other if the distance
between them is smaller than $1/\kappa$.  At these distance scales the
gas will therefore not be dilute at all.  To wit, in this case the
chemical potential $\mu_\sigma$ defined by eq. (\ref{fugb}) is large
and negative.

Another situation in which the dilute gas approximation is valid,
is for those variational functions $D(p)$ which have a constant value
$D(p)=1/2\kappa$ for all momenta larger than some scale $\mu$, such
that $\mu a<<1$. In this case $\mu_\sigma$ is close to zero, and one
is back to the case of pure Coulomb gas.
For these variational functions the behaviour of the electric vortex
gas is very similar to the behaviour of the magnetic vortex gas. Both
have BKT phase transition at $n=8$, both are in the molecular phase
for $n>8$ and in the plasma phase at $n<8$.

The question is however, is there any reason to expect, that the best
variational propagator will behave in this fashion.
Surprising as it may seem at the first glance, the answer to this is positive.
In fact, a little thought convinces one that
it is almost unavoidable in a compact theory with
finite radius of compactness $1/g$.  The point is the following. As
discussed in Section 2, imposing compact gauge invariance, among other
things has an effect of imposing the following condition on the
magnetic field
\begin{equation}
e^{ia^2gnB(x)}|\Psi>=|\Psi>
\label{Bn}
\end{equation}
This means, that a gauge invariant state $|\Psi>$ has nonzero
projection only on states with quantized eigenvalues of the magnetic
field $B$: $B(x)=\frac{2\pi}{n}\frac{a^{-2}}{g}$. We know, however
that in the noncompact vacuum $<B^2>\propto a^{-3}$. This means that
the natural scale for the magnetic field in the vacuum state of the
noncompact theory is $B\propto a^{-3/2}$. Therefore if one projects a
state with the ultraviolet properties of the noncompact vacuum onto a
gauge invariant state, only the contributions of $B=0$ states will
survive and the magnetic energy will vanish. Since the
expectation value of the magnetic field in a noncompact state is just
given by $<B^2>=\int \frac{d^2p}{4\pi^2}p^2D(p)$, a nontrivial compact
dynamics can be described only by states whith a much larger value of
the ``propagator'' $D(p)$. In fact, since $D(p)$ is bounded from
above, nontrivial fluctuations of magnetic field can survive only if
$D(p)=O(1/\kappa)$ for all momenta between the ultraviolet cutoff
$a^{-1}$ and some intermediate scale $\mu$, which itself is much less
than the cutoff. In this case one has $<B^2>\propto a^{-4}\kappa^{-1}$
and the scale of the fluctuations is just right to satisfy
eq. (\ref{Bn}) in a nontrivial way.

We see therefore that the emergence of the intermediate scale $\mu$ is
mandatory in the compact theory, and the dilute gas approximation for
electric vortices gas should be reliable. It turns out however, that
one can not determine the value of the scale $\mu$ within the dilute
approximation itself. The reason is that, at least in the molecular
phase, the behaviour of the electric vortices is very insensitive to
the exact value of the crossover scale $\mu$, as long as it is much
smaller than the ultraviolet cutoff. This is so because in the Coulomb
gas, away from the critical point the vortices are bound in pairs of
the characteristic size of order of the ultraviolet cutoff
$a$. Therefore if the interaction potential is changed at distances
$x>1/\mu>>a$, the behaviour of the gas hardly changes at all.

Even
though the precise value of $\mu$ can not be determined, we can
establish reliably the important fact that $\mu>>\kappa$. This is of
course crucial, since if this was not the case the propagator would be
constant practically for all momenta, and the theory would have no
propagating degrees of freedom in the continuum limit.  The reason
this can be established, is that, as will be shown in the next section
within the dilute gas approximation the scale $\mu$ is pushed up all
the way to the ultraviolet cutoff. Of course, as discussed earlier,
one cannot believe the approximation for such large $\mu$. However,
the dilute gas approximation is expected to be valid for values of
$\mu$ which are much larger than $\kappa$ as long as they are much
smaller than the ultraviolet cutoff.

To verify this picture and also estimate the value of the crossover
scale $\mu$ we will also perform a calculation using a different
approximation for the electric vortex gas. This approximation is valid
for the propagator functions $D(p)$ which vanish at large momentum, so
that functions with the ultraviolet asymptotics of the noncompact
propagator can be studied reliably.  To define this approximation we
rewrite the partition function of the electric gas in terms of the
integer valued field $\tau$ defined in eq. (\ref{tau}) as
\begin{equation}
Z_\sigma=\int D\tau D\beta\exp\left\{-\pi n\kappa\tau p^2
D\tau\ +i\int d^2x
2\beta(x)(\cos\tau(x)-1)\right\}
\end{equation}
Approximating the Lagrange multiplier field $\beta$ by a constant (the
same kind of approximation which lead to eq. (\ref{sine}) for magnetic
vortices) gives a tractable expression.
\begin{equation}
Z_\sigma=\int D\tau\exp\left\{-\pi n\kappa\tau p^2
D\tau\ +2a^{-2}\int d^2x(\cos\tau(x)-1)\right\}
\end{equation}

It is again convenient to define an appropriate fugacity $\xi$ in
 analogy to (\ref{zeta}).
\begin{equation}
\frac{n}{4\pi}\xi=a^{-2}<cos(\tau)>=
a^{-2}\exp\left\{-\frac{\pi}{n}\int \frac{d^2p}{(2\pi)^2}
\left[\kappa p^2D(p)+\xi\right]^{-1}\right\}
\label{xi}
\end{equation}
With the noncompact function $D$, this fugacity is exponentially small
\begin{equation}
\xi\propto \exp(-\frac{r}{n}(a\kappa)^{-1})
\end{equation}
where $r$ is a numerical constant.  For this type of variational
functions therefore, electric vortices can be consistently treated in
this ``$\tau$ - approximation''.

\section{Minimization of the energy and the phases of the model.}

\subsection{ The expectation value of the energy.}

Now we are ready to compute the expectation value of the Hamiltonian
eqs. (\ref{HB},\ref{HE}). We start with the magnetic part eq.(\ref{HB}).
Straightforward integration over the vector potential and the
noncompact part of the gauge group gives
\begin{eqnarray}
<\exp\{i2ga^2B(x)\}>=\exp\left\{-a^4g^2\int
\frac{d^2p}{4\pi^2}p^2D(p)\right\}\times \\ \nonumber
<\exp\{i2a^2\pi\rho(x)\}>_\rho
<\exp\left\{-2a^2\kappa
\int \frac{d^2p}{4\pi^2}e^{ipx}D(p)\sigma(p)\right\}>_\sigma
\label{magneticenergy}
\end{eqnarray}
Where $<>_\rho$ and $<>_\sigma$ mean average over the magnetic and
electric vortex ensembles correspondingly. The magnetic vortex
contribution is trivial, since $a^2 \rho(x)$ is an integer.

We calculate the electric vortex contribution in the dilute gas
approximation, keeping the first order correction in $\zeta$. A
straightforward calculation gives
\begin{eqnarray}
& &<\exp\left\{-2a^2\kappa
\int \frac{d^2p}{4\pi^2}e^{ipx}D(p)\sigma(p)\right\}>_\sigma
= \\ \nonumber
& &\exp\left\{
\frac{8\pi\kappa}{n}a^4\zeta\int \frac{d^2p}{4\pi^2}
\frac{p^2D(p)}{\frac{1}{2\pi n\kappa}p^2D^{-1}(p)+2\zeta}
+2\zeta\int d^2 x \left[\cosh X(x)-1
-\frac{1}{2}X(x)\right]
 \right\}
\end{eqnarray}
where the function $X(x)$ is defined by
\begin{equation}
X(x)=\frac{2}{n}a^2\int
\frac{d^2p}{4\pi^2}e^{ipx}\frac{p^2}{\frac{1}{2\pi
n\kappa}p^2D^{-1}(p)+2\zeta}
\label{X}
\end{equation}

An analogous calculation for the electric energy part, eq. (\ref{HE})
gives
\begin{eqnarray}
& &<\exp\{\frac{i2}{n}\alpha_i(x)E_i(x)\}> =  \exp\left\{-\frac{\pi
a^2}{2n\kappa}\int \frac{d^2p}{4\pi^2}(D^{-1}(p)+4\kappa^2
D(p))\right\}  \\ \nonumber
& & <\exp\left\{-\frac{\pi}{\kappa}\int
\frac{d^2p}{4\pi^2}e^{ipx}\rho(p)
p^{-2}D^{-1}(p)\epsilon(p)\right\}>_\rho ~
 <\exp\left\{-\frac{\kappa}{\pi}\int
\frac{d^2p}{4\pi^2}e^{ipx}\sigma(p)D(p)\omega(p)\right\}>_\sigma
\label{electricenergy}
\end{eqnarray}

where we have defined
\begin{equation}
\epsilon=\frac{g}{2\pi}\epsilon_{ij}\partial_i\alpha_j; \ \ \
\partial^2\omega=g\partial_i\alpha_i
\end{equation}

The electric vortex contribution is calculated as

\begin{equation}
\exp\left\{\frac{4\pi\kappa a^2}{n}\zeta\int
\frac{d^2p}{4\pi^2}\frac{D(p)}{\frac{1}{2\pi n\kappa}
p^2D^{-1}(p)+\zeta} +2\zeta\int d^2x \left[\cosh
Y(x)-1-\frac{1}{2}Y^2(x)\right]\right\}
\end{equation}
with
\begin{equation}
Y(x)=\frac{2}{n}\int
\frac{d^2p}{4\pi^2}e^{ipx}\frac{p^2\omega(p)}{\frac{1}{2\pi
n\kappa}p^2D^{-1}(p)+2\zeta}
\label{Y}
\end{equation}

The magnetic vortex contribution this time is nontrivial and is given by

\begin{equation}
\exp\left\{\frac{\pi a^2}{\kappa n}z\int \frac{d^2p}{4\pi^2}
{}~\frac{D^{-1}(p)}{\frac{2\kappa}{\pi n} p^2D(p)+2z}+2z\int d^2x\left[\cosh
L(x)-1-\frac{1}{2}L^2(x)\right]\right\}
\end{equation}
with
\begin{equation}
L(x)= \frac{2}{n }~\int \frac{d^2p}{4\pi^2}e^{ipx}\frac{\epsilon (p)}{
\frac{2\kappa}{\pi n}p^2D(p)+2z}
\label{L}
\end{equation}}

Here we also give the expressions for the electric vortex
contributions calculated in the ``$\tau$''-approximation:
\begin{eqnarray}
& &<\exp\left\{-2a^2\kappa
\int \frac{d^2p}{4\pi^2}e^{ipx}D(p)p^2\tau(p)\right\}>_\sigma =
 \\ \nonumber
& &\exp\left\{a^4g^2\int
\frac{d^2p}{4\pi^2}\frac{\kappa D^2(p)p^4}{\kappa p^2 D(p)+\xi}+
\frac{n\xi}{2\pi}\int d^2 x \left[\cos I(x)-1-\frac{1}{2}I^2(x)\right]
\right\}
\label{magneticenergy1}
\end{eqnarray}
and
\begin{eqnarray}
& &<\exp\left\{-\frac{\kappa}{\pi}\int
\frac{d^2p}{4\pi^2}e^{ipx}\tau(p) p^2 D(p)\omega(p)\right\}>_\sigma= \\
\nonumber
& &\exp\left\{\frac{2\pi a^2}{n}\int
\frac{d^2p}{4\pi^2}\frac{\kappa^2 p^2 D^2(p)}{\kappa p^2D(p)+\xi}
+\frac{2n}{4\pi}\xi\int d^2x \left[\cos
J(x)-1+\frac{1}{2}J^2(x)\right]\right\}
\end{eqnarray}

with $I(x)$ and $J(x)$ defined by
\begin{equation}
I(x)=\frac{4\pi a^2\kappa}{n}
\int \frac{d^2p}{4\pi^2}e^{ipx}\frac{p^2
D(p)}{\kappa p^2D(p)+\xi}
\label{I}
\end{equation}
\begin{equation}
J(x)=\frac{2}{n}\int \frac{d^2p}{4\pi^2}e^{ipx}\frac{\kappa p^2
D(p)\omega(p)}{\kappa p^2D(p)+\xi}
\label{J}
\end{equation}

We will now use these expressions to minimize the average
 energy.  We will consider  the cases $n > 8$ and $n < 8$
 separately.

\subsection{$n > 8$}
 As was discussed in the previous section, for $n > 8$ the magnetic
 vortex gas is in the dipole phase. The vortex fugacity $z$ vanishes,
 and in the dilute gas approximation they do not contribute to the
 energy.

The electric vortex gas will also be treated here in the
 dilute gas approximation. As explained in the previous section, the
 minimization here is performed on the set of variational functions
 $D(p)$ which have the constant value $D(p)=1/2\kappa$ for all momenta
 above some scale $\mu<<a^{-1}$. For these variational functions, the
 electric vortex gas is indeed dilute, and for $n>8$ is in the
 molecular phase. The electris fugacity $\zeta$ also vanishes, and
 therefore the electric vortices also do not contribute to the energy.

The calculation is further simplified, since as can be checked
explicitly with these variational functions $D(p)$ the expectation
values in eqs.(\ref{magneticenergy}) and (\ref{electricenergy}) for
large $n$ are close to one.  Therefore the relevant exponentials can
be expanded up to first order in Taylor series.

In this approximation, therefore the expectation value of energy is
the same as for the noncompact theory and the minimization with
respect to $D(p)$ gives
\begin{equation}
D^{-1}(p)=\sqrt{p^2+4\kappa^2}, \ \ \ p^2<\mu^2
\end{equation}
We therefore find that for $n>8$ both magnetic and electric vortices
are irrelevant, and the compact theory at small momenta is
indistinguishible from the noncompact one.

One assumption that was made in the previous discussion is that the value of
the propagator $D$ at large momenta is exactly $1/2\kappa$. The
general argument given in the previous section pretty much establishes
that this value should be of order $1/\kappa$, but does not establish
the coefficient.  This magic number $1/2$ enabled us to conclude that
the critical value of $n$ for the electric vortex gas is the same as
for magnetic vortex gas, $n=8$. To verify this picture further, we
would like to demonstrate explicitly that above the scale $\mu$ this
is indeed the correct behaviour of the propagator.
To this end we will minimize the energy calculated in the
``$\tau$-approximation''. This calculation is valid as long as the
fugacity $\xi$ is small. To make sure this condition is satisfied, we
will first consider only such functions $D(p)$ that behave
$D(p)\propto1/\sqrt{p^2}$ for momenta $\lambda^2<p^2<a^{-2}$ with
$\lambda<a^{-1}$.  Note, that although $\lambda$ must be smaller than
the ultraviolet cutoff, it does not have to be parametrically smaller,
so that for example $\lambda a$ may remain a finite small constant in
the limit $a\rightarrow 0$.

The terms depending on $I$ and $J$ (see (\ref{J}), ~(\ref{L})) in the
 expression for the expectation value of the energy can be neglected
 at large $n$, since they give corrections supressed by powers of
 $1/n^2$.  Now, substituting the formulae of the previous section into
 equations (\ref{HB}) and (\ref{HE}) we obtain for the expectation
 value of the energy $H$
\begin{equation}
2V^{-1}<H> = \frac{1}{2}~\int~\frac{d^2p}{(2\pi)^2}
\left[D^{-1}(p) + (p^2 + 4\kappa^2) D(p)
\frac{\xi}{\kappa p^2 D(p) + \xi} \right]
\label{exp}
\end{equation}

Minimizing this with respect
to $D(p)$ we obtain
\begin{equation}
D^{-2}(p)=Z(p^2+M^2)\frac{\xi^2}{(\kappa p^2 D(p)+\xi)^2}
\label{minim}
\end{equation}
where the constants $Z$ and $M$ are
\begin{equation}
Z=1+\frac{\kappa}{\xi}\frac{\partial \ln \xi}{\partial \ln
D}\frac{(\kappa p^2 D+\xi)^2}{\kappa p^2 D}\int
\frac{d^2p}{4\pi^2}(p^2+4\kappa^2)\frac{\kappa p^2D^2(p)}{(\kappa
p^2D(p)+\xi)^2}
\label{Z}
\end{equation}
\begin{equation}
M^2=4\kappa^2Z^{-1}
\label{M}
\end{equation}
The solution of eq.(\ref{minim}) is
\begin{equation}
D^{-1}(p)=Z^{1/2}(p^2+M^2)^{1/2}-\frac{\kappa}{\xi}p^2
\label{solution}
\end{equation}
Since $D^{-1}$ is bounded from below, this solution is valid for
momenta for which $D^{-1}>2\kappa$, that is for
\begin{equation}
p^2<\mu^2=\frac{4\xi^2}{M^2}(1-\frac{M^2}{\xi})
\label{mu}
\end{equation}
For momenta greater than $\mu$ and smaller than $\lambda$
\begin{equation}
D^{-1}(p^2>\mu)=2\kappa
\end{equation}

We see that the physics corresponding to this solution is precisely
the same as we were describing earlier. Even though we have not
determined the value of the scale $\mu$ yet, let us for the moment
assume that it satisfies the relation $M<<\mu<<a^{-1}$.  Then at low
momenta $p^2<<\mu^2$ the variational propagator is the same as in the
noncompact case. Therefore the physics it describes at all physical
scales is the same as in the noncompact theory.  The only difference
is that there appears a nontrivial mass and wave function
renormalization given by the same factor $Z$ in eqs.(\ref{Z},\ref{M}),
which is small as long as $\xi$ is small enough.  At scales $p^2>\mu$
the propagator is frozen, and therefore at these momenta the theory
does not contain propagating degrees of freedom. Thus effectively, the
ultraviolet cutoff has been also changed from $a^{-1}$ to a lower
value $\mu$. Note also that for $\xi<<\lambda^2$, which is where the
approximation is valid, the scale $\mu$ also satisfies
$\mu<<\lambda$. The emergence of the scale $\mu$ and the change in the
behaviour of the propagator at large momenta is therefore seen already
in the regime where the $\tau$-approximation is valid.  It is again
difficult to establish the precise value of this scale $\mu$.  The
point is that it grows very fast when the auxilliary scale $\lambda$
approaches the ultraviolet cutoff.
It is clear from eqs.(\ref{solution},\ref{mu}) that the dynamics
favours precisely this situation.
However, when $\lambda$ is close to
$a^{-1}$, the $\tau$-approximation ceases to be valid.

A rough estimate of $\mu$ can be obtained by
pursuing the $\tau$-approximation  to the end.
Calculating the integral in (\ref{xi}) using  an explicit
 expression for $D(p)$ (\ref{solution}) one has after some algebra
\begin{equation}
\xi = \frac{4\pi}{n} a^{-2} \exp\left(
-\frac{1}{3n} \frac{\xi}{M^2} -
\frac{1}{n} \ln \frac{M}{2\xi a}
\right)
\end{equation}
Using the fact that $\ln (1/Ma) >>1$ the expression for fugacity
is as follows
\begin{equation}
\xi = 3n M^2 \ln \frac{4\pi}{ na^2 M^2}
\left[1 + O\left(\ln\ln(1/Ma)/\ln(1/Ma)\right)\right]
\label{xi1}
 \end{equation}

{}From this we get
\begin{equation}
\mu= 6n M \ln \frac{4\pi}{ na^2 M^2}
\label{mu1}
\end{equation}
We see indeed, that $\mu/M\rightarrow_{a\rightarrow 0}\infty$.

We stress again, that the exact value of $\mu$ as given in
eq.(\ref{mu1}) is not to be trusted literally. It is obtained in the
region of variational functions $D(p)$ where
the nonlinearities in the effective $\tau$
theory are not at all small. For example, there is a large
(logarithmic in cutoff) renormalization of the fugacity $\xi$ due to
the cosine interaction already at the two loop level. On the other
hand we believe that this renormalization will be the leading
effect. If it is taken into account properly, the expressions
(\ref{solution}, \ref{xi1}, \ref{mu1}) will not be changed drastically
if $\xi$ appearing in them is understood as the complete renormalized
fugacity.  We have checked for example, that the $I$ and $J$ dependent
terms in the energy expectation value, which appear due to the
nonlinearity in the $\tau$ action, are supressed relative to the terms
we have kept in eq.(\ref{exp}) even for fugacities $\xi$ of
eq.(\ref{xi1}).  We believe therefore that the estimate for the
crossover scale $\mu$ eq.(\ref{mu1}) is qualitatively correct.

To summarize this part, we find that for $n>8$ the compactness of the
theory is not important for infrared physics. The best variational
propagator at physical momenta is identical to the propagator in the
noncompact theory. The only effect of the finite radius of compactness,
 $1/g$ is that it ``freezes'' the propagation of the modes
with high momentum, thereby effectively just changing the
 ultraviolet cutoff.

\subsection{$n<8$.}

Let us now turn to a more interesting case $n<8$. At these values of
$n$ both, magnetic and electric vortices are in the plasma phase. The
fugacities $z$ and $\zeta$ do not vanish, and in the dilute gas
approximation we obtain for the expectation value of the energy

\begin{eqnarray}
2V^{-1}<H> &=& \frac{1}{2}~\int~\frac{d^2p}{(2\pi)^2}
D^{-1}(p)\left[1-\frac{2z}{\frac{2\kappa}{\pi n}p^2D(p)+2z}\right]
 \nonumber  \\ \nonumber
& +& (p^2 + 4\kappa^2) D(p)\left[1- \frac{2\zeta}{\frac{1}{2\pi
n\kappa} p^2 D^{-1}(p) +2\zeta} \right] \\
&-&\frac{n}{2\pi
\kappa}a^{-4}\zeta\int d^2x (\cosh X-1-\frac{1}{2}X^2) \\ \nonumber
&-&\frac{4\kappa n}{\pi}a^{-2}\zeta\int d^2x (\cosh
Y-1-\frac{1}{2}Y^2) \\ \nonumber
&-&\frac{4\kappa n}{\pi}a^{-2}z\int
d^2x (\cosh L-1-\frac{1}{2}L^2)
\label{nlesseight}
\end{eqnarray}

The last three terms appear due to nonlinearities in the $\psi$ and
$\chi$ Lagrangians. They are neglible as long as the appropriate gases
are dilute, that is $\zeta$ and $z$ are smaller than the ultraviolet
cutoff. We present these terms here for completeness, but will neglect them
in the following derivations. It can be checked that on the solution we
obtain, these terms are indeed small.

Minimizing the energy with respect to $D(p)$ we obtain after some algebra
the following equation
\begin{eqnarray}
\left(\frac{2\kappa}{\pi n}p^2D(p)+2z\right)^2\left[\frac{1}{2\pi n
\kappa}p^2 (p^2+4\kappa^2)+A\zeta\right]= \\ \nonumber
\left(\frac{1}{2\pi n\kappa}
p^2 +2\zeta D(p)\right)^24\kappa^2\left[\frac{2\kappa}{\pi
n}p^2+Bz\right]
\end{eqnarray}
with constants
\begin{equation}
A=\frac{1}{16\pi^4n\kappa}\left[1-\zeta\int \frac{d^2 q}{4\pi^2}\frac{1}
{(\frac{1}{2\pi
n\kappa} q^2 D^{-1}(q) +2\zeta)^2}\right]^{-1}
\int \frac{d^2 q}{4\pi^2}\frac{q^2(q^2+4\kappa^2)}
{(\frac{1}{2\pi
n\kappa} q^2 D^{-1}(q) +2\zeta)^2}
\label{A}
\end{equation}
and
\begin{equation}
B=\frac{4\kappa}{8\pi^3 n}
\left[1-z\int \frac{d^2 q}{4\pi^2}\frac{1}{
(\frac{2\kappa}{\pi n} q^2 D(q) +2z)^2}\right]^{-1}
\int \frac{d^2 q}{4\pi^2}\frac{q^2}{
(\frac{2\kappa}{\pi n} q^2 D(q) +2z)^2}
\label{B}
\end{equation}

The solution of this quadratic equation is
\begin{equation}
D(p)=\left\{\frac{p^2}{4\pi n\kappa}[\zeta F(p)-4\kappa^2
z]-(z\zeta-\frac{p^4}{4\pi^2 n^2})\sqrt{F(p)}\right\}\left[\zeta^2
F(p)-\frac{\kappa^2 p^4}{\pi^2 n^2}\right]^{-1}
\label{notgood}
\end{equation}
where we have defined
\begin{equation}
F(p)=\frac{4\kappa^2[\frac{2\kappa}{\pi n}p^2+Bz]}{\frac{1}{2\pi n
 \kappa}p^2 (p^2+4\kappa^2)+Az}
\label{F}
\end{equation}

Let us analyze these expressions. First, note that both values of fugacities
are of the order of the fugacity of the Coulomb gas. This is true, since
as we already know from the previous analysis for $p^2>\mu$, the variational
propagator is a constant $1/2\kappa$. Then the integrals in the definitions
of the chemical potentials
$\mu_\rho$ and $\mu_\sigma$ in (\ref{fuga}) and (\ref{fugb})
get contributions only from momenta lower than $\mu<<a^{-1}$.

The fugacity of the Coulomb gas behaves like
\begin{equation}
a^{-2}e^{-\frac{8 c}{8-n}}
\end{equation}
Therefore for $n\ne 8$ it is smaller, but of the order of the
ultraviolet cutoff $a^{-2}$.
We can then estimate the constants $A$ and $B$ as
\begin{equation}
A\propto \frac{a^{-2}}{\kappa}, \ \ \ \ B\propto\kappa
\end{equation}
This gives
\begin{equation}
F(0)\propto\frac{\kappa^4 z}{a^{-2}\zeta}
\end{equation}
Substituting this into the expression in eq.(\ref{notgood}) we find
\begin{equation}
D(0)=\frac{z}{\zeta \sqrt {F(0)}}\propto
\sqrt{\frac{z}{\zeta}}\frac{a^{-1}}{\kappa}\frac{1}{\kappa}>>\frac{1}{2\kappa}
\end{equation}
 It is also easy to see that the function $D(p)$ as given in
eq. (\ref{notgood}) is a growing function at small momenta, and is of
the same order of magnitude up to momenta of order $p^2\sim \kappa
a^{-1}$.  On the other hand the variational function $D(p)$ by
definition, can not exceed the value $1/2\kappa$.  It follows therefore
that the solution eq. (\ref{notgood}) is unphysical and we have to
choose for $D(p)$ the end point value
\begin{equation}
D(p)=\frac{1}{2\kappa}
\end{equation}
at {\it all momenta}.

We therefore discover that for $n<8$ the propagator is constant for
all momenta, or is completely local in the coordinate space. This
should remind one of the behaviour of correlation functions in
topological theories: there correlation functions of any local
operator are completely local.  The situation therefore is such that
for $n<8$ the Maxwell term in our theory becomes totally irrelevant,
and the theory degenerates into a pure Chern - Simons theory.  This
conclusion is further confirmed by inspection of the expectation value
of the energy (\ref{nlesseight}). Note, that the structure of this
equation is such that for momenta which are smaller than $z$ and
$\zeta$ the contributions of the electric and magnetic vortices
cancell completelly the ``noncompact'' contributions.  Since on our
solution $\zeta$ and $z$ are both of the order of the ultraviolet
cutoff\footnote{Note that with this function $D(p)$ $z$ and $\zeta$ are
{\it both} equal to the fugacity of the Coulomb gas.}, physical
momenta do not contribute at all to the energy.  This is again
consistent with a topological theory, in which the Hamiltonian
vanishes.

\subsection{Spontaneous breaking of magnetic flux at $n<8$.}

Let us discuss in more detail the physical properties of the theory
for $n<8$. As we have just argued, it has many similarities with a
pure topological theory. It is clear, however, that it can not be
equivalent to a simple noncompact pure Chern - Simons theory. First,
as is obvious from the previous calculation, compactness of the gauge
group is crucial in this phase. It is responsible for the appearance
of the magnetic and electric vortex gases, which are in the plasma
phase and thereby determine the infrared properties of the
model. Second, our derivation in the previous subsection is in large
measure independent of the assumption $\kappa<<a^{-1}$. For large
$\kappa=O(a^{-1})$, the ``photon mass'' is of the order of the
ultraviolet cutoff, and this case corresponds to the pure Chern -
Simons limit of the TMGT. It is clear therefore, that the compact pure
Chern - Simons theory also undergoes phase transition at $n=8$, the
small $n$ phase being qualitatively different from the large $n$
phase.

We conclude that the compact TMGT at $n<8$ is equivalent to
compact pure Chern - Simons model, but they both must be different
from the noncompact Chern - Simons theory.  What distinguishes the two
phases of the pure Chern - Simons theory?

The fundamental property of a noncompact Chern - Simons theory is the
Bohm - Aharonov interaction between charged particles. It is
interesting to check, therefore whether this interaction is still
there in the compact theory for $n<8$.  A straightforward way to study
this question is to calculate the expectation value of the Wilson loop
in a state which contains a unit external charged. The Bohm - Aharonov
interaction should show up as the Bohm - Aharonov phase in this
expectation value.  Introduction of a unit external charge at the
point $x=0$ leads to the following modification of the Gauss' constraint
equation
\begin{equation}
\partial_iE_i(x)-2\kappa B(x)=g\delta^2(x)
\end{equation}
The ground state waive functional in this sector should be well
approximated by a projected Gaussian with a nonzero shift in the
vector potential.  For simplicity, we will take the width of the
Gaussian to be the same as in the zero charge sector
$G^{-1}(p)=\kappa$.
\begin{equation}
|1>=\int Ds_iD\phi\exp\left\{i\kappa
\left[(\epsilon_{ij}(s_i-\partial_i\phi)A_j
+\epsilon_{ij}\partial_i\phi s_j\right]-ig\phi(0)
-\frac{\kappa}{2} (A_i^{\phi,s}-a_i)^2\right\}
\end{equation}
The function $a_i$ should be treated as a variational function.
Since we are dealing with the pure Chern - Simons theory, the function
$a_i$ should be holomorphic
\begin{equation}
a_1+ia_2=0
\end{equation}
Additional constraint on $a_i$ follows from the requirement
that the state $|1>$ be normalizable. This forces $a_i$ to
satisfy at zero momentum
\begin{equation}
{\rm Re}~ \epsilon_{ij}\partial_ia_j+{\rm Im}~ \partial_ia_i=
-\frac{g}{\kappa}\delta^2(x)
\label{normal}
\end{equation}

We will see
that the result does not depend on the detailed form of $a_i$.
The expectation value of the
Wilson loop in this state can be calculated in a straightforward manner.
\begin{equation}
<\exp(ig\int_SBdS)>=\exp(ig\int_S\epsilon_{ij}\partial_i a_j dS)
<\exp\left(\frac{g}{2}\int_S \partial^2\phi dS\right)>_\phi
<\exp\left(-\pi\int_S\sigma
dS\right)>_\sigma
\end{equation}
where for holomorphic functions $a_i$, the weight
for the $\sigma$ averaging is the same as in the vacuum state,
while the weight
for the $\phi$ averaging is given by
\begin{equation}
\exp\{-S[\phi]\}=
\exp\left\{-\frac{\kappa}{2}(a_i+\partial_i\phi)^2+\frac{\kappa}{4}
(a_i+a^*_i+\partial_i\phi+i\epsilon_{ij}\partial_j\phi)^2+
ig\phi(0)\right\}
\end{equation}
Finally we obtain
\begin{equation}
<\exp\left(ig\int_SBdS\right)>=R(S)~
  \exp\left(-i\frac{2\pi}{n}\right)
\end{equation}
where $R(S)$ is a real number. The value of $R(S)$ depends on the
phase of the theory, but the value of the phase does not. Also, $R(S)$
is nonzero in both phases.

We conclude therefore that the Bohm - Aharonov interaction
 of external charges is the same in the compact theory
for $n>8$ and $n<8$.

Nevertheless, the two phases are distinguishible. The operator that
distinguishes them is not the Wilson loop, but is closely related to
its dual. It is the 2+1 dimensional analog of the t'Hooft loop
\cite{thooft} - the operator that creates magnetic vortices.  The
significance of this operator is, that it is an order parameter for
spontaneous breaking of the magnetic flux symmetry \cite{kr}.

In fact, the fate of the magnetic flux symmetry in the compact TMGT
is an interesting question.
Recall, that in noncompact electrodynamics in 2+1 dimensions
(both with and
without the Chern - Simons term) the homogeneous Maxwell equation
\begin{equation}
\partial_\mu\tilde F_\mu=0
\end{equation}
ensures the existance of the conserved current. The global charge
associated with this current is the magnetic flux through the plain
\begin{equation}
\Phi=\int d^2x B(x)
\end{equation}
In QED without Chern - Simons term, this charge is spontaneously
 broken in the
Coulomb phase. This breaking is accompanied by the appearance
of the massless Goldstone boson - the photon.
In noncompact TMGT the photon is massive. Accordingly the magnetic
flux is
not broken, but annihilates the vacuum state.
In a compact theory the magnetic flux is not conserved anymore
\cite{kr}.
The magnetic ``monopole - instantons'' change the magnetic flux
 through the
plain by an integer multiple of $2\pi/g$. Consequently, only the
 following
subgroup of the flux group remains the symmetry of the theory
\begin{equation}
U_N=e^{gN\Phi}
\end{equation}
for integer $N$.  In QED without Chern - Simons term, the magnetic
flux is not quantized, and therefore the operators $U_N$ constitute
the group of integers $Z$. This symmetry again is spontaneously
broken.  Since the group is discreet, its spontaneous breakdown does
not require an existence of a massless particle, and the photon in
compact QED is massive.  In TMGT the situation is slightly
different. Recall, that the compactification of the gauge group in
TMGT requires the magnetic flux $\Phi$ to be quantized in units of
$2\pi/ng$. On these states only the operators $U_N$ for $N=1,... ,
n-1$ are represented nontrivially. The magnetic flux symmetry in
compact TMGT is therefore just $Z_n$.  Let us note,
 that i t is this reduction of
the magnetic flux group to $Z_n$ that is responsible for the mixing of
the states with $n/2$ positively and negatively charged particles in
the spontaneously broken SU(2) TMGT discussed in  paper \cite{lee}.

The natural question is, what is the realization of this $Z_n$ flux
symmetry in compact TMGT. To answer this question we may calculate the
vacuum expectation value of the appropriate order parameter.  This
order parameter should be an operator which is gauge invariant under
both, compact and noncompact gauge groups, and should transform
nontrivially under the flux symmetry. The suitable operator is
\begin{equation}
v_m(x,C)=\exp\left\{i\frac{2\pi m}{gn}\int_C
\epsilon_{ij}dx_iE_j\right\}
\end{equation}
The contour $C$ is a semiinfinite line with an endpoint at $x$.  For
integer $m<n$ this operator commutes with the elements of the compact
gauge group.  The calculation of the expectation value of $v$ is
straightforward and parallels exactly the calculation of the electric
part of the expectation value of the energy.
Without giving the details here, we just describe the result.

For $n>8$,
there are no vortex contributions and the result is basically the same as in
the noncompact theory. It has the form
\begin{equation}
<v_m>=\exp\{-Ka^{-1}L\}
\end{equation}
where $K$ is a numerical constant, and $L$ is the length of the curve
$C$. In the infinite volume, $L\rightarrow\infty$, and we find
\begin{equation}
<v>=0, \ \ \ n>8
\end{equation}

For $n<8$ the contributions of the magnetic and electric vortices
cancell exactly the noncompact contributions, precisely in the same way as they
did in the calculation of the energy expectation value. There is therefore
no linear divergence in the infinite volume limit, and we obtain
\begin{equation}
<v>\ne 0,\ \ \ n<8
\end{equation}

We conclude therefore, that for $n>8$, the $Z_n$ magnetic flux group in
compact TMGT is unbroken, and for $n<8$ it is broken spontaneously.
These different ways the symmetry is realized in the vacuum
distinguish  between the two phases of the model.

\section{Discussion.}

To summarize, the result of our variational calculation is the
following.  For $n>8$ the magnetic and electric vortices in the wave
function are bound in pairs. Their only effect is to introduce an
intermediate scale $\mu$, below which the physics is the same as in
the noncompact theory. This scale $\mu$ becomes infinite in the limit
of infinte ultraviolet cutoff, and therefore in the continuum limit
the effects of compactness disappear.  This result is in qualitative
agreement with the results of previous studies \cite{CSmonopole}.
Here we want to make the following remark. The basic picture of
ref. \cite{CSmonopole} is that the monopoles in the compact Chern -
Simons theory are bound in pairs by a {\it linear} potential. On the
other hand the magnetic (and electric) vortices in our ground state
wave functional have interaction which is logarithmic at large
distances. It is important to realize that these two claims are not
inconsistent with each other. The point is, that the object which in
path integral formalizm is represented by a monopole - antimonopole
pair has a segment of a Dirac string, stretched between them.  This
Dirac string becomes ``observable'' in the Chern - Simons theory and
carries a finite action density, which is the origin of the linear
potential between the monopoles. In our Hamiltonian description, a
widely separated monopole - antimonopole pair is represented by a
configuration in the wave function which has a nonvanishing magnetic
vortex density $\rho$ at two points, but also a string of electric
vortex dipoles $\sigma$ along the line that connects these
points. This line of electric vortex dipoles corresponds precisely to
the ``observable'' segment of the Dirac string. Configurations of this
type are indeed suppressed in our wave functional by an exponential of
the length of the dipole string segment. In this sence it is indeed
true, that monopoles in our wave functional interact with a linear
potential.

The new result we find, is that for $n<8$ the nature of the ground
state of the theory is very different. Due to the liberation of
magnetic and electric vortices, the correlation functions of local
observables (such as $B$ or $E$) become completely local. The theory
therefore does not describe any propagating degrees of freedom.  The
liberation of the vortices also leads to spontaneous breaking of the
magnetic flux symmetry in this phase. In this respect this phase is
similar to the confining compact QED without the Chern -
Simons term, whose vacuum also spontaneously breaks magnetic flux.

The dichotomy between the ``monopoles'' on one hand and the
Chern-Simons term on the other hand, is resolved therefore in this
nontrivial way.

Finally, let us remark, that in recent years TMGT have seen many
applications in condensed matter physics, both in relation to Quantum
Hall Effect and high temperature superconductivity. In the latter case
a prominent role is played by the so-called semions. Those are charged
particles coupled to Chern-Simons theory with $n=4$.  It is clear from
our results, that the question of compactness of the Chern-Simons
theory must be crucial for the ``semion physics''. The physics usually
discussed, corresponds to the noncompact theory. In the compact case
the semions must have a very different behaviour. We hope to return to
this question in future work. \\

\noindent
{\bf Acknowledgements.}
One of us (I.K.) is  grateful to A.I. Vainshtein, M.A. Shifman
 and all members of TPI for many  interesting discussions and hospitality
 during the visits in  March-April and Summer of 1995.
 A.K. would like to thank members of Theoretical Physics Department
 at Oxford for hospitality during his visit in the Fall 1994, when
 this project has been started.
We are grateful to A. Bochkarev for many useful discussions.
This work was supported in part by DOE under grant number DE-FG02-94ER40823,
PPARC grant GR/J 21354 and by Balliol College, University of Oxford.

\newpage

{\renewcommand{\Large}{\normalsize}

\end{document}